\documentclass[12pt]{iopart}
\usepackage{iopams} 
\usepackage{graphicx}
\usepackage{amssymb}
\usepackage{epstopdf}

\begin{document}

\title{Multi-band spectroscopy of inhomogeneous Mott-insulator states of ultracold bosons}

\author{D.~Cl\'ement, N.~Fabbri, L.~Fallani, C.~Fort and M.~Inguscio}
\address{LENS, Dipartimento di Fisica,
Universit\`a di Firenze and INFM-CNR, via Nello Carrara 1, I-50019 Sesto Fiorentino (FI), Italy.}
\ead{clement@lens.unifi.it}
\begin{abstract}
In this work, we use inelastic scattering of light to study the response of inhomogeneous Mott-insulator gases to external excitations. The experimental setup and procedure to probe the atomic Mott states are presented in detail. We discuss the link between the energy absorbed by the gases and accessible experimental parameters as well as the linearity of the response to the scattering of light. We investigate the excitations of the system in multiple energy bands and a band-mapping technique allows us to identify band and momentum of the excited atoms. In addition the momentum distribution in the Mott states which is spread over the entire first Brillouin zone enables us to reconstruct the dispersion relation in the high energy bands using a single Bragg excitation with a fixed momentum transfer.
\end{abstract}

\submitto{\NJP}

\maketitle
\tableofcontents

\section{Introduction}
 
Many-body quantum systems where strong correlations between particles play a central role are among the most intriguing physical systems since no simple picture captures their behaviour. Examples of this kind can be found in correlated electronic systems such as high-Tc superconductors \cite{DagottoRMP1994}, quantum liquids such as liquid Helium \cite{NoziereBook} or interacting 1D systems \cite{GiamarchiBook} as carbon nanotubes \cite{IshiiNature2003}, quantum wires \cite{AuslaenderScience2002} and organic conductors \cite{SchwartzPRB1998}. A large amount of research work has been devoted to the study of such systems within different fields of physics, yet many questions have still to be addressed. In this prospect new experimental possibilities have been opened by the realization of degenerate quantum gases and the recent development of technics to manipulate them in optical lattices. Indeed, in the past years ultra-cold atomic gases have been loaded in optical lattices to create strongly correlated quantum phases in a highly controllable manner. 

Among these experiments, different atomic insulating phases have been realized demonstrating the versatility of gaseous systems. These include bosonic Mott insulators in one-dimensional (1D), two-dimensional (2D) and three-dimensional (3D) systems \cite{StoferlePRL2004,SpielmanPRL2007,GreinerNature2002}, disordered bosonic insulating phases \cite{FallaniPRL2007} as well as fermionic Mott insulators \cite{JordensNature2008,SchneiderScience2008}. The characterization of these insulating quantum phases has enlightened many of their properties. Experiments have demonstrated the presence of a gap in the spectrum of Mott insulators \cite{StoferlePRL2004,GreinerNature2002} and its vanishing in a disordered insulating phase \cite{FallaniPRL2007}. A noise correlation analysis of time-of-flight pictures has shown the spatial order of the atomic distribution \cite{FollingNature2005} and its controlled modification when modifying the lattice potential \cite{GuarreraPRL2008}. Studies of the suppression of compressibility has been performed in fermionic \cite{JordensNature2008,SchneiderScience2008} and bosonic \cite{LignierPRA2009} Mott insulators. In addition to these properties expected in homogeneous insulating phases, the presence of a trap enriches the experimental situation while creating alternate regions of Mott with different filling factors and superfluid states \cite{JakschPRL1998,BatrouniPRL2002}. This shell structure has been identified in experiments and clearly related to the trapping potential \cite{CampbellScience2006,FollingPRL2006}. Recently, inelastic light scattering has been used to measure the excitation spectrum of inhomogeneous Mott insulating states \cite{ClementPRL2009}. This spectroscopic measurement performed at non-zero momentum transfer and in the linear response regime gives a direct access to the dynamical structure factor $S$ of those complex phases \cite{OostenPRA2005,ReyPRA2005,PupilloPRA2006,KollathPRL2006,HuberPRB2007,MenottiPRB2008}.

In this paper we report the measurement of the excitation spectrum of 1D bosonic Mott insulating states over several energy bands. The band structure is induced by the presence of the optical lattice which drives the system into a Mott insulating state. A particular attention is paid to the details of the spectroscopic technic. We discuss the measurement of the energy absorbed by the atomic system when excitations are created as well as the linearity of its response. This paper also extends our previous work \cite{ClementPRL2009} the scope of which was centered on the study of excitations lying in the lowest energy band. We present experimental results for the spectra in the higher energy bands showing novel experimental signatures related to the one-particle spectral function of the Mott state. A band-mapping technic is used to identify the momentum of the excited atoms contributing to the different parts of the spectra in the inhomogeneous Mott state. 

The paper is organized as in the following. In section \ref{InelasticScatt} we describe the principles and the experimental implementation of inelastic light scattering (Bragg spectroscopy) as a probe of excitations of ultra-cold atoms in optical lattices. In section \ref{EnergyLinResp} we present the experimental sequence allowing to extract information about the amount of excitation created by the Bragg beams. We discuss the choice of the experimental parameters in order to work in the linear regime, in particular for the response of Mott insulating states. Section \ref{MultiBandSpect} is devoted to the multi-band spectra of the inhomogeneous Mott insulating states and to the band-mapping technic. The latter allows us to directly construct the dispersion relation in the high energy bands induced by the optical lattice. Finally we discuss the relation of the high-energy band spectra with the measurement of the one-particle spectral function of the atomic Mott state.

\section{Inelastic light scattering from correlated gases in optical lattices}\label{InelasticScatt}

\subsection{Inelastic light scattering as a probe of excitations}

The result of weak inelastic scattering of waves or particles by many-body systems may be described within the Born approximation and expressed in terms of the dynamic structure factor $S$ \cite{VanHovePR1954}. The dynamic structure factor is the Fourier transform of the density-density correlation function and it carries information on the excitation spectrum of the system, independently from the external probe. The knowledge of the dynamic structure factor $S$ obtained from inelastic scattering processes has proved to be central in describing many-body systems, from the electrons in solids \cite{AshcroftBook} to superfluid Helium \cite{NoziereBook}.

Inelastic light scattering has been applied to gaseous Bose-Einstein condensates to measure the dynamic structure factor \cite{KozumaPRL1999,StengerPRL1999,DavidsonRMP}. This scattering technic, referred to as Bragg spectroscopy, consists in a two-photon transition between two different momentum states of the same internal ground-state \cite{KozumaPRL1999}. The dispersion relation of interacting BECs in the mean-field regime \cite{StamperKurnPRL1999,Steinhauer2002,Steinhauer2003}, the presence of phase fluctuations in elongated BECs \cite{Richard2003} as well as signatures of vortices \cite{Muniz2006} have been investigated using this technic. Bragg spectroscopy is also used as a tool to coherently manipulate atomic clouds for interferometric schemes \cite{MullerPRL2008} or for thermodynamics studies \cite{GerbierPRA2004,InadaPRL2008}. More recently it has succeeded in providing novel information about strongly interacting 3D Bose \cite{Papp2008} and Fermi \cite{Veeravalli2008} gases close to Feshbach resonances as well as correlated 1D Bose gases across the transition from the superfluid to the Mott insulator state \cite{ClementPRL2009}.

\subsection{Two-photon Bragg transitions}

\begin{figure}[ht!]
\begin{center}
\includegraphics[width=0.6\columnwidth]{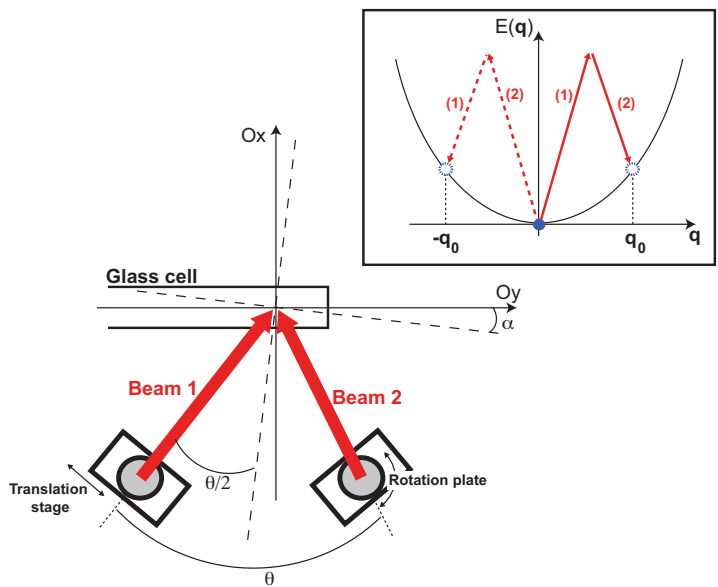}
\end{center}
\caption{Experimental setup: two laser beams (red), detuned with respect to each other by a controllable frequency $\nu$, are shone onto the atoms in order to induce a two-photon transition. The direction of each beam can be changed thanks to a rotation plate and a translation stage that allow a precise control of the angle $\theta$. The axes $(Ox,Oy)$ coincide with the axes of the lattice beam (see text below). Inset: parabolic energy spectrum of free particles. Depending on the sign of the relative detuning $\nu$ between the Bragg beams atoms absorb photons from beam $1$ and emit in beam $2$ (for $\nu>0$, solid lines) or absorb from beam $2$ and emit in beam $1$ (for $\nu<0$, dashed lines), changing the direction of the momentum transfer $\hbar \textbf{q}_{0}$.}
\label{Fig1}
\end{figure}

The Bragg spectroscopy technic consists in shining the atoms for a finite time $t_{B}$ with two laser beams detuned by $\delta$ from an atomic transition. During the Bragg pulse, atoms absorb photons from one beam and are stimulated to emit photons in the second beam. The resonance condition for the two-photon transition consists in conserving energy and momentum. An atom with an initial momentum $\textbf{p}_{i}$ ends up in the same internal state with a final momentum $\textbf{p}_{f}=\textbf{p}_{i}+ \hbar \textbf{q}_{0}$ where $\hbar \textbf{q}_{0}$ is the momentum transfer given by the two-photon process. The energy difference between the initial and final atomic states is also given by the Bragg beams, which are detuned from each other by a frequency $\nu=\nu_{1}-\nu_{2}$ where $\nu_{1}$ and $\nu_{2}$ are the frequencies of the two beams (see Fig.~\ref{Fig1}). In practice, by tuning the frequency difference $\nu$, one can match the resonance energy for the two-photon transition process to occur at a given momentum transfer $\hbar \textbf{q}_{0}$. In other words, this corresponds to the condition for which the atomic wave-function is efficiently diffracted on the moving optical lattice created by the Bragg beams. We note $V_{B}$ the amplitude of the lattice induced by the Bragg beams and $\Omega_{B}/2 \pi=V_{B}/2 h$ the associated Rabi frequency.
 
The modulus of the momentum transfer $\hbar \textbf{q}_{0}$ given to the atoms is set by the wavelength $\lambda_{B}$ of the Bragg beams and the angle $\theta$ between them, namely $\mathrm{q}_{0}=4 \pi / \lambda_{B} \sin (\theta/2)$ (see Fig.~\ref{Fig1}). Note that the direction of $\textbf{q}_{0}$ is perpendicular to the bisector between the two beams. The detuning $\nu$ between the two Bragg beams controls not only the energy $h\nu$ transferred to the atoms but also the sign of the momentum transfer. For a two-photon transition towards an excited state with higher kinetic energy, atoms absorb a photon from the beam with the higher-energy photons and are stimulated to emit a photon in the beam with the lower-energy photons. Therefore, with our convention, when the detuning $\nu=\nu_{1}-\nu_{2}$ is positive atoms absorb photons from beam $1$, emit into beam $2$ and the momentum transfer is $+\hbar \mathrm{q}_{0}$ (see inset in Fig.~\ref{Fig1}). When the detuning $\nu$ is negative atoms absorb from beam $2$, emit into beam $1$ and the momentum transfer is opposite, \textit{i.e.} $-\hbar \mathrm{q}_{0}$. 

In our experimental setup, a laser beam coming out from a laser diode at wavelength $\lambda_{B}=780~$nm, detuned by $\delta \simeq 300~$GHz from the D2 transition of $^{87}$Rb, is separated into two beams, each passing through an Acousto-Optic Modulator (AOM). The two AOMs are locked in phase and allow to control the relative detuning $\nu$ ($\nu=\nu_{1}-\nu_{2}\ll \delta$) between the two Bragg beams. After passing through the AOMs, each beam is injected into a polarization-maintaining fiber. The out-coupler of each fiber is fixed onto a post-holder mounted on a rotation plate and a translation stage close to the cell in which the atoms are manipulated. This setup allows an independent control of the angle of each of the Bragg beams with respect to the y-axis (see Fig.~\ref{Fig1}). In particular the angle $\theta$ between the two Bragg beams can be tuned from $30^{\circ}$ to $60^{\circ}$, this range being determined by the limited optical access to the glass cell. While $\lambda_{B}$ is precisely known in the experiment, the angle $\theta$ cannot be measured geometrically with high precision. We use the atoms as a sensor to precisely calibrate $\textbf{q}_{0}$. To this purpose we measure the momentum transferred to a 3D BEC (in the absence of any optical lattice) in two different ways. 

\subsection{Calibration of the momentum transfer}

In the first calibration measurement, we measure the resonant frequency of the Bragg spectrum for the 3D BEC. The resonance frequency depends on the momentum transfer $\hbar \textbf{q}_{0}$, on the strengh of atom-atom interactions through the dispersion relation and on the initial velocity of the BEC center of mass. The dispersion relation $E(\textbf{q})$ of a trapped 3D BEC in the mean-field regime can be written as
\begin{equation}
E(\textbf{q})=\hbar \sqrt{c^2_{\mathrm{LDA}}(\textbf{q}) \textbf{q}^2 + \left ( \frac{\hbar \textbf{q}^2}{2m} \right)^2}
\end{equation}
where $c_{\mathrm{LDA}}(\textbf{q})$ corresponds to an effective sound velocity within a Local Density Approximation and is related to the mean-field interaction term \cite{ZambelliPRA2000}. The two-photon transition induced by the Bragg pulse couples the initial state of the 3D BEC with momentum $\textbf{p}_{i}=\hbar \textbf{q}_{i}$ to an excited state with momentum $\textbf{p}_{f}=\hbar (\textbf{q}_{i}+\textbf{q}_{0})$. Thus the resonance energy of this process is $E(\textbf{q}_{i}+\textbf{q}_{0})-E(\textbf{q}_{i})$. In the absence of interactions, this energy reduces to the usual quadratic dependence of a single-particle spectrum, \textit{i.e.} $\hbar^2(\textbf{q}_{i}+\textbf{q}_{0})^2/2m-\hbar^2\textbf{q}_{i}^2/2m$. In order to reduce the effect on evaluating $\textbf{q}_{0}$ coming from interactions we perform Bragg spectroscopy on dilute 3D BECs after a time-of-flight. When the magnetic trap is switched off the BEC acquires a spurious non-zero momentum $\hbar \textbf{q}_{i}$ and two unknown parameters have to be determined, namely $\textbf{q}_{0}$ and $\textbf{q}_{i}$. We measure the spectra at positive ($\hbar q_{0}$) and negative ($-\hbar q_{0}$) momentum transfer. The two resonance frequencies $E(\textbf{q}_{i}+\textbf{q}_{0})-E(\textbf{q}_{i})$ and $E(\textbf{q}_{i}-\textbf{q}_{0})-E(\textbf{q}_{i})$ allow us to precisely determine $\rm{q}_{0}$ in the experiment. We obtain $\rm{q}_{0}=0.97(3) k_{L}$ where $k_{L}=2 \pi /\lambda_{L}$ is the wave-vector of the lattice beams (see text below) at the wavelength $\lambda_{L}=830~$nm.

In the experiment, the bisector of the two Bragg beams is not exactly perpendicular to the y-axis and the momentum transfer along the y-axis is the projection $\hbar \mathrm{q}_{0,y}=4 \hbar \pi \cos (\alpha) / \lambda_{B} \sin (\theta/2)$ where $\alpha$ is the angle as defined on Fig.~\ref{Fig1}. Note that the axes on Fig.~\ref{Fig1} are defined as the axes of the lattice beams (see below). To measure the angle $\alpha$ we diffract the atoms in the Raman-Nath regime \cite{OvchinnikovPRL1999} with very short light pulses (of typical duration $3~\mu$s). In this diffraction regime, many clouds of diffracted atoms are observed allowing a good estimate of the axis of the momentum transferred to the atoms. We use such a procedure for both the lattice beams and the Bragg beams separately, measuring the angle $\alpha=9.5(1)^{\circ}$ (see Fig.~\ref{Fig1}). As a consequence the projection of the momentum transfer along the y-axis is $\hbar \mathrm{q}_{0,y}=0.96(3) \hbar k_{L}$.

In a second set of calibration measurements, we use the diffracted atoms of the 3D BEC by the moving lattice created with the Bragg beams. By letting the atoms fall under gravity after the Bragg pulse for a long enough time-of-flight ($t_{\mathrm{TOF}}=10-30~$ms), the diffracted atoms separate from the atoms which have not undergone the two-photon transition. The distance between the two clouds is $\hbar \mathrm{q}_{0,y} t_{\rm{TOF}} /m$ in the $yOz$ plane where absorption images are taken, $m$ being the atomic mass. By fitting the distance between the diffracted and non-diffracted atomic clouds as a function of $t_{\mathrm{TOF}}$, we measure $\mathrm{q}_{0,y}=0.97(4) k_{L}$ in good agreement with the previous measurement. 

\section{Amount of excitation and linear response regime} \label{EnergyLinResp}

\subsection{Energy transfer by the Bragg beams to a gas loaded in optical lattices}

\subsubsection{Experimental setup and time sequence}

We use the Bragg spectroscopic scheme described in section \ref{InelasticScatt} to probe the response of correlated 1D Bose gases in the linear response regime \cite{ClementPRL2009}. Correlated Bose phases are created by loading a 3D BEC of $^{87}$Rb atoms in a 3D optical lattice as described in previous papers \cite{FallaniPRL2007,ClementPRL2009}. In brief, a 3D BEC of $N\sim1.5 \times 10^5$ atoms is produced in a magnetic trap which frequencies are $\omega_{x}=\omega_{z}=2\pi \times 90~$Hz and $\omega_{y}=2\pi \times 8.9~$Hz. Then it is adiabatically loaded in a 3D optical lattice created with three pairs of counter-propagating laser beams at the wavelength $\lambda_{L}=830~$nm. The amplitudes $V_{i}$ of the optical lattice along each axis $i=x,y,z$ are expressed in units of the recoil energy $E_{R}=h^2/2 m \lambda_{L}^2$, $V_{i}=s_{i}E_{R}$. The optical lattices are ramped up to their final values $s_{i}$ with an exponential ramp of duration $140~$ms and time constant $30~$ms. 

In all the experiments described in this paper, a 2D optical lattice in the plane $xOz$ with amplitudes $s_{x}=s_{z}=s_{\perp}=35$ is used to create an array of 1D tubes of atoms. At this amplitude $s_{\perp}$ and in the absence of the longitudinal lattice ($s_{y}=0$), we measure the ratio of the frequencies of the quadrupole mode $\nu_{Q}$ to the dipole mode $\nu_{D}$ to check that the 1D gases are Bose-condensed. For holding times in the 2D lattice (at $s_{\perp}=35$) up to $50~$ms, this ratio is close to the value $\sqrt{3}$ corresponding to the expected value for 1D Bose-condensed gases close to the mean-field regime \cite{MenottiPRA2002,MoritzPRL2003}. For longer holding times, the ratio increases up to the value $2$ expected for a thermal 1D Bose gas. All the experiments are performed with a typical holding time at the amplitude $s_{\perp}=35$ equal to $10~$ms ensuring the 1D gases are Bose-condensed in the absence of the longitudinal lattice ($s_{y}=0$). 

The amplitude $s_{y}$ of the optical lattice is changed to tune the 1D gases from a correlated superfluid state ($s_{y}=0$) to inhomogeneous Mott insulating states ($s_{y}>6$). For large amplitudes $s_{y}$ the dynamics of 1D gases (when restricted to the lowest energy band of the optical lattice along $Oy$) can be described by the Bose-Hubbard Hamiltonian \cite{JakschPRL1998},
\begin{equation}
H=-J \sum_{<i,j>} (a_{i}^{\dag}a_{j}+ \mathrm{h.c.}) + \frac{U}{2} \sum_{i} n_{i} (n_{i}-1) + \sum_{i} \epsilon_{i} n_{i}.\label{eqBHH}
\end{equation}
Here $a_{i}^{\dag}$, $a_{i}$ are the creation and annihilation operator of one boson at site $i$ and $n_{i}=a_{i}^{\dag}a_{i}$ is the particle number operator. The on-site interaction energy is given by $U$ and the next-neighbour hopping amplitude by $J$. $\epsilon_{i}$ is the slowly-varying energy offset experienced by an atom on site $i$ due to the presence of the harmonic confining potential.

Once the amplitude of the 3D optical lattice has reached its final values ($s_{\perp}=35,s_{y}$) the Bragg pulse is shone onto the system for a time $t_{B}$ (see Fig.~\ref{Fig2}(a)). In order to detect the amount of excitation induced by the Bragg beams in the correlated gaseous systems we follow a procedure similar to that used in \cite{StoferlePRL2004} and described in \cite{ClementPRL2009} that consists in measuring a quantity related to the increase of energy in the gas.

\subsubsection{Measurement of the energy transfer}

\begin{figure}[ht!]
\begin{center}
\includegraphics[width=\columnwidth]{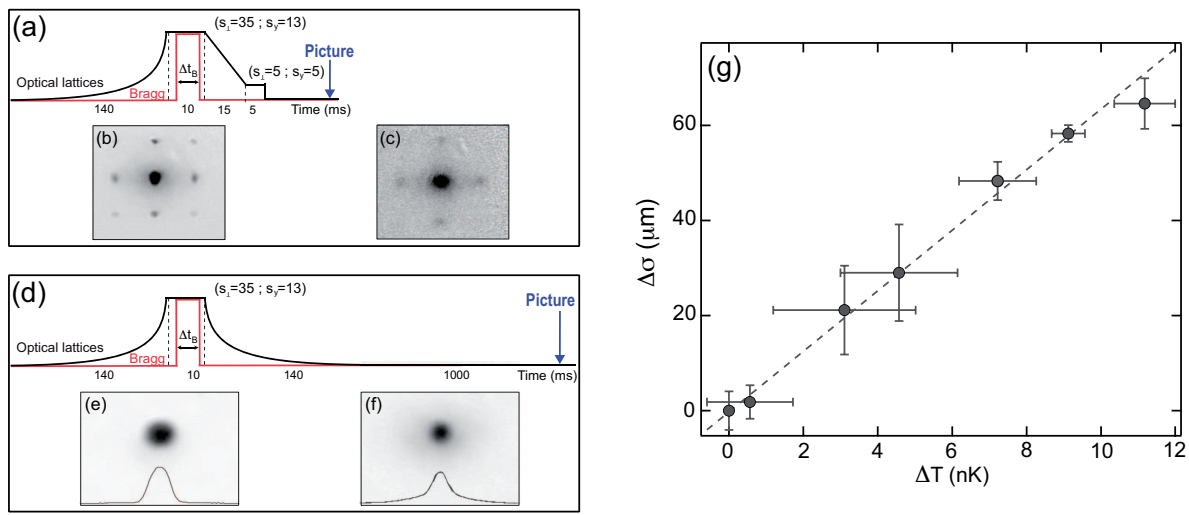}
\end{center}
\caption{(a) Experimental sequence to measure the increase $\Delta \sigma$ of the width of the central peak of the interference pattern after a Bragg pulse. The picture (b) [\emph{resp.} (c)] corresponds to an absorption image taken after a non-resonant (\emph{resp.} resonant) Bragg pulse. (d) Experimental sequence to measure the increase of temperature $\Delta T$ after a Bragg pulse and a thermalization time of $1~$second. The picture (e) [\emph{resp.} (f)] corresponds to an absorption image taken after a non-resonant (\emph{resp.} resonant) Bragg pulse. (g) $\Delta \sigma$ as a function of the increase of temperature $\Delta T$. The points correspond to different power of the Bragg beams with a fixed pulse length of 3 ms.}
\label{Fig2}
\end{figure}

After shining the Bragg pulse onto the atoms the amplitudes of the 3D lattice are ramped down linearly to $s_{\perp}=s_{y}=5$ in $15~$ms where we let the system rethermalize for $5~$ms [see Fig.~\ref{Fig2}(a)]. Then the magnetic trap and the 3D lattice are switched off abruptly and an absorption image of the atomic distribution in the $yOz$ plane is taken after a typical time-of-flight $t_{\mathrm{TOF}}\simeq22~$ms. Expanding from a phase coherent state in a 3D optical lattice ($s_{\perp}=s_{y}=5$), the atomic distribution exhibits an interference pattern which is the analogous of the diffraction pattern of light from a grating (see Fig.~\ref{Fig2}(b) and \cite{GreinerPRL2001}). From this interference pattern we extract the RMS widths $\sigma_{y}$ and $\sigma_{z}$ of the central peak, the increases of which are related to that of the energy of the system. Indeed, when the Bragg excitation is tuned out of resonance both widths $\sigma_{y}$ and $\sigma_{z}$ are equal to the widths measured in the absence of the Bragg pulse while on resonance they increase [see Fig.~\ref{Fig2}(c)]. Moreover we find that $\sigma_{y}$ and $\sigma_{z}$ have the same dependence with the detuning $\nu$ between the Bragg beams as expected from an efficient re-thermalization process in each spatial direction when the lattices are ramped down. In the following we will consider an average width $\sigma$ defined as $\sigma=\sqrt{\sigma_{y}\sigma_{z}}$. 

To be more quantitative on the relation between the increase of $\sigma$, noted $\Delta \sigma$, and the energy absorbed by the gas, we make a comparison of $\Delta \sigma$ after a resonant Bragg excitation with a measurement of the temperature increase $\Delta T$ realized in the same experimental conditions (in particular for an identical Bragg excitation). The temperature is obtained by ramping down adiabatically the optical lattices with an exponential ramp after the Bragg pulse, letting the system thermalize in the magnetic trap for 1~s and measuring the condensate fraction of the 3D atomic cloud [see Fig.~\ref{Fig2}(d)]. The comparison is performed by setting the detuning $\nu$ on the resonance towards the second band in the case of an inhomogeneous Mott insulator state with $s_{y}=13$ (see Section~\ref{MultiBandSpect}). By changing the power in the Bragg beams, $\Delta \sigma$ is tuned over a range typical of that used in experiments. The results are presented in Fig.~\ref{Fig2} where $\Delta T$ is extracted from the decrease of the condensate fraction. This measurement confirms that $\Delta \sigma$ is proportional to the energy absorbed by the atomic system, \textit{i.e.} to the number of excitations created. 

One may wonder whether the spectra could be obtained measuring the condensate fraction instead of the width $\sigma$. This could be the case when the amplitude of the response of the 1D gases is large as in the superfluid regime or in the Mott regime for high-energy bands. On the contrary, the amplitude of the response of the Mott insulating state within the lowest energy band is low as the system exhibits its insulating behaviour. In this case, we observe that the measurement of $\Delta \sigma$ is more sensitive than that of the condensate fraction, allowing to detect small excitations.

\subsection{Amount of excitation $\mathcal{A}$}

We now describe the quantity we will be referring to as the amount of excitations. This quantity is proportional to the increase $\Delta \sigma$ of the central peak width of the interference pattern obtained after a time-of-flight (see Fig.~\ref{Fig2}(b) and text below). It is rescaled as explained in the following.

\begin{figure}[ht!]
\begin{center}
\includegraphics[width=0.5\columnwidth]{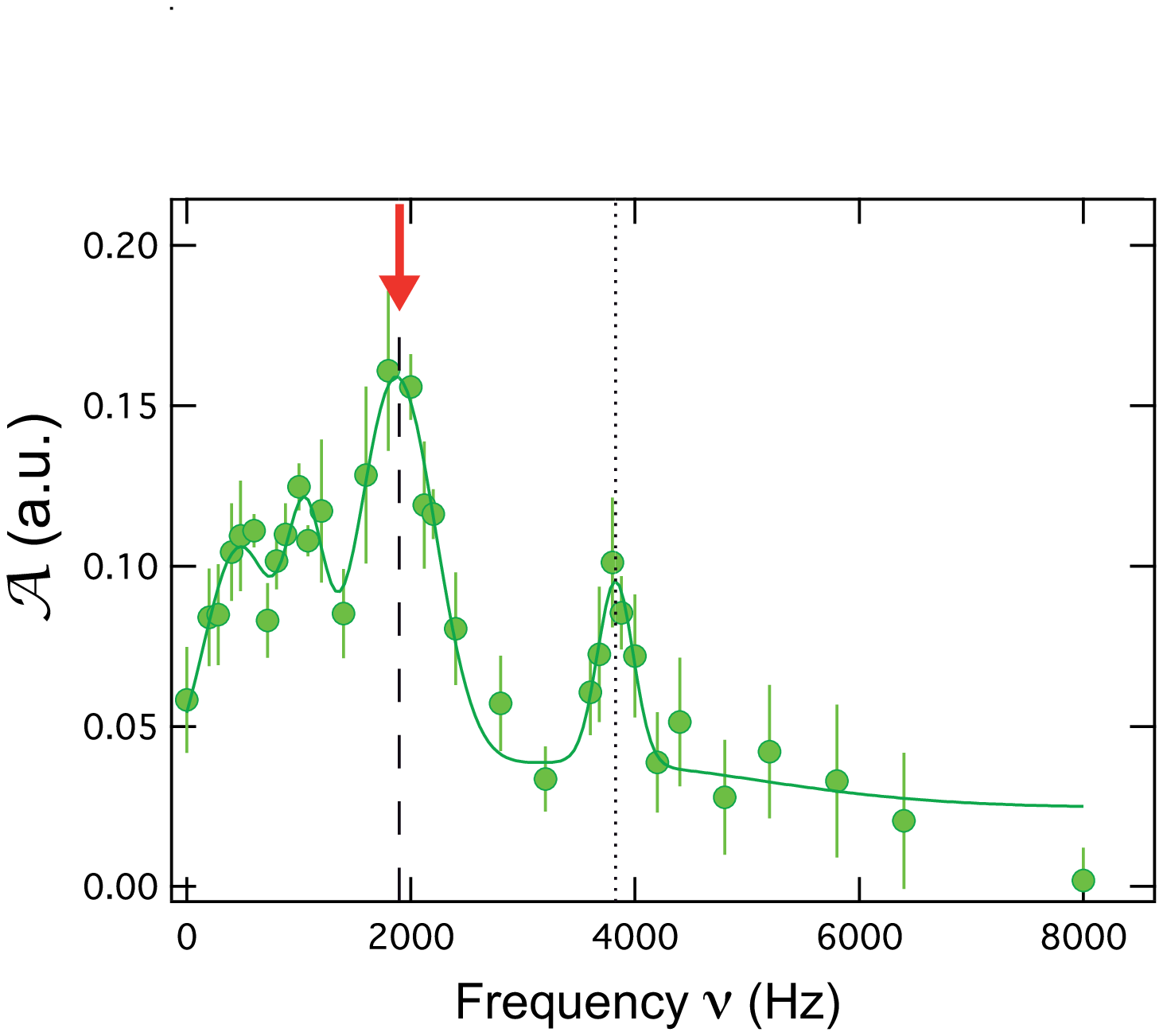}
\end{center}
\caption{Bragg spectrum of an inhomogeneous Mott insulating state with lattice amplitudes $s_{\perp}=35$ and $s_{y}=13$. The dashed (\textit{resp.} dotted) vertical line marks the resonance energy $\Delta_{\mathrm{ph}} \sim U$ (\textit{resp.} $2\Delta_{\mathrm{ph}} \sim 2U$) of the Mott state. The red arrow indicates the detuning $\nu$ at which the linearity of the response has been tested. It corresponds to the resonance energy of a particle-hole excitation in the Mott regions.}
\label{Fig3}
\end{figure}

The width $\sigma$ of the central peak has a weak dependence on the total atom number $N$ which is related to the presence of interactions, independently from the Bragg excitation. In the experiment, the total atom number typically fluctuates by $20\%$ from shot-to-shot. In order to remove the small variations (up to $8\%$) on the measurement of $\sigma$ coming from fluctuations of $N$, we calibrate the increase of $\sigma$ with $N$ by changing the atom number over a wide range $10^5-10^6$. This scaling has been used to substract, for each measurement of $\sigma$, the contribution to $\sigma$ coming from the fluctuations in the atom number. To display the spectrum the width in the absence of Bragg excitation is substracted in order to plot the increase $\Delta \sigma$. In addition, we rescale $\Delta \sigma$ by the parameters of the Bragg excitation, namely by the factor $(P_{B} t_{B}/\delta)^{-1}$ where  $P_{B}$ is the power of the Bragg beams, $\delta$ their detuning with respect to the atomic transition and $t_{B}$ the duration of the Bragg pulse \cite{NoteScaling}. This procedure allows to compare the amplitudes of the response for spectra taken with slightly different parameters of the Bragg beams. The linear dependence of $\Delta \sigma$ with $(P_{B} t_{B}/\delta)^{-1}$ holds in the linear response regime, the validity of which we discuss below. In the following we will refer to the rescaled $\Delta \sigma$, called $\mathcal{A}$, as to the amount of excitation transferred to the system.  The plot of $\mathcal{A}$ as a function of the detuning $\nu$ between the two Bragg beams gives the excitation spectrum. In Fig.~\ref{Fig3} we show an example of a spectrum of the inhomogeneous Mott insulating state (see text below and  \cite{ClementPRL2009}).

\subsection{Linear response regime}

\begin{figure}[ht!]
\begin{center}
\includegraphics[width=0.9\columnwidth]{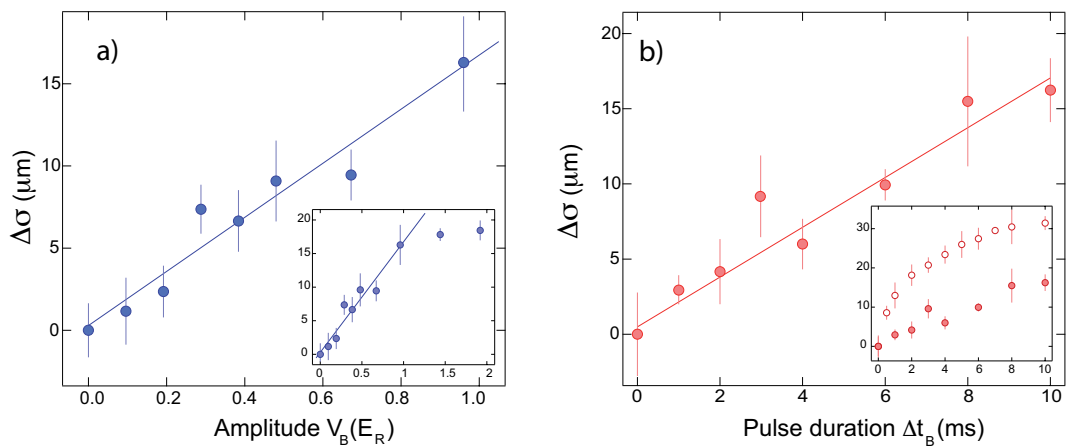}
\end{center}
\caption{(a) $\Delta \sigma$ as a function of the amplitude $V_{B}$ of the lattice induced by the Bragg beams which is proportional to the power $P_{B}$. The measurement is performed at a fixed energy transfer $\nu=\Delta_{\mathrm{ph}}/h$ corresponding to the resonance of the particle-hole excitation and for a duration of the Bragg pulse $\Delta t_{B}=3~$ ms. Inset: same measurement on larger scale of the amplitude $V_{B}$ showing saturation for $V_{B}>E_{\rm{R}}$. (b) $\Delta \sigma$ as a function of the duration of the Bragg pulse at a given amplitude $V_{B}=0.5 E_{\rm{R}}$. As in (a), the measurement is performed at the energy $\nu=\Delta_{\mathrm{ph}}/h$ of the particle-hole excitation. Inset: comparison of the response for two amplitudes $V_{B}$, namely $V_{B}=0.5 E_{\rm{R}}$ (dots) and $V_{B}=2.0 E_{\rm{R}}$ (circles).}
\label{Fig4}
\end{figure}

The linearity of the response of the system to the Bragg excitation is studied monitoring how $\Delta \sigma$ varies with the parameters of the Bragg excitation on resonance. In \cite{ClementPRL2009} we have identified the particle-hole excitation energy $\Delta_{\mathrm{ph}}$ of the Mott insulating state. Here we fix the frequency $\nu=\Delta_{\mathrm{ph}}/h$ (\textit{i.e.} on the resonant peak of the Mott insulating state, see red arrow on Fig.~\ref{Fig3}) to study the variations of $\Delta \sigma$ as a function of the amplitude $V_{B}$ of the Bragg lattice and the duration $\Delta t_{B}$ of the pulse, with the results plotted on Fig.~\ref{Fig4}. In the range of parameters used to monitor the spectra (for example $V_{B}= 0.5 E_{\rm{R}}$ and $\Delta t_{B}=6~$ms for the spectrum in Fig.~\ref{Fig3} at $s_{y}=13$), the increase of $\Delta \sigma$ is linear both with $V_{B}$ and $\Delta t_{B}$, demonstrating that the energy transferred to the gases by the Bragg pulse increases linearly with both those quantities. 

The inset of Fig.~\ref{Fig4}b) compares the variation of $\Delta \sigma$ as a function of the pulse duration $\Delta t_{B}$ for two amplitudes $V_{B}$, namely $V_{B}=0.5 E_{\rm{R}}$ (dots) and $V_{B}=2.0 E_{\rm{R}}$ (circles) at $\nu=\Delta_{\mathrm{ph}}/h$ for $s_{y}=13$. For the larger amplitude $V_{B}=2.0 E_{\rm{R}}$ the response to the Bragg excitation exhibits a saturation and the time scale of the linear regime is shorter. We note that we did not observe Rabi oscillations as in the case of an excitation from the ground-state of a 3D BEC in a harmonic trap and a Bogoliubov mode \cite{DavidsonRMP}. In the latter case the excited state is decoupled from the ground-state, a situation which might not be true anymore in the Mott state where strong correlations are present. The increase of $\Delta \sigma$ also saturates for large amplitudes $V_{B}$ as depicted in the inset of Fig.~\ref{Fig4}a).

In addition, we stress that the amplitude $V_{B}$ of the moving lattice created by the Bragg beams for such parameters is much smaller than the amplitude $s_{y}$ of the longitudinal lattice: all experiments are performed in a regime where $V_{B} < 0.05 V_{y}$. This regime is different from the one where the Bragg spectrum has been obtained using the lattice modulation technic. In the latter case, the amplitude of the lattice is typically modulated by $\sim 30\%$ due to a momentum transfer $\mathrm{q}_{0}=0$ \cite{StoferlePRL2004,FallaniPRL2007}. In this work, the parameters $U$ and $J_{y}$ (see Eq.~\ref{eqBHH}) describing the system in the Bose-Hubbard model are almost unaltered by the additional light potential created by the Bragg beams.

The probe we use (Bragg beams) weakly affects the state of the system under investigation and the amount of excitation varies linearly with the parameters of the probe. In this linear response regime the spectra we measure in the lowest energy band are proportional to the dynamical structure factor $S$ \cite{VanHovePR1954}.

\section{Excitations in inhomogeneous Mott states towards different energy bands}\label{MultiBandSpect}

Bragg spectroscopy has been used in \cite{ClementPRL2009} to investigate the response of inhomogeneous Mott insulating states within the lowest energy band and to study their properties on an energy scale of the order of the Mott gap $\Delta_{\mathrm{ph}}$. Excitations to higher energy bands have been recently reported in \cite{MullerPRL2007} where the authors focused on the lifetime and the coherence of higher bands population. Here, we study the spectrum of inhomogeneous Mott insulating states on a large energy scale, focusing on transitions to high-energy bands of the optical lattice, for a momentum transferred by the Bragg beams $\mathrm{q}_{0,y}=0.96(3) k_{L}$. 

Since the typical energy transferred for excitations in high energy bands is at least one order of magnitude larger than the Mott gap (see details later), the excited atoms are not correlated to the non-excited ones lying in the lowest energy band. In addition, the Bragg excitation being weak, only a small fraction of the atomic cloud is excited. Therefore, the small amount of atoms excited in the high energy bands can safely be considered as non-interacting particles. This point makes the study of the high-energy excitation spectrum of a Mott insulating state much different from that restricted to the lowest energy band where atom-atom correlations play a crucial role \cite{ClementPRL2009}.

\subsection{Multi-band spectrum of a Mott insulating state}

\begin{figure}[ht!]
\begin{center}
\includegraphics[width=0.75\columnwidth]{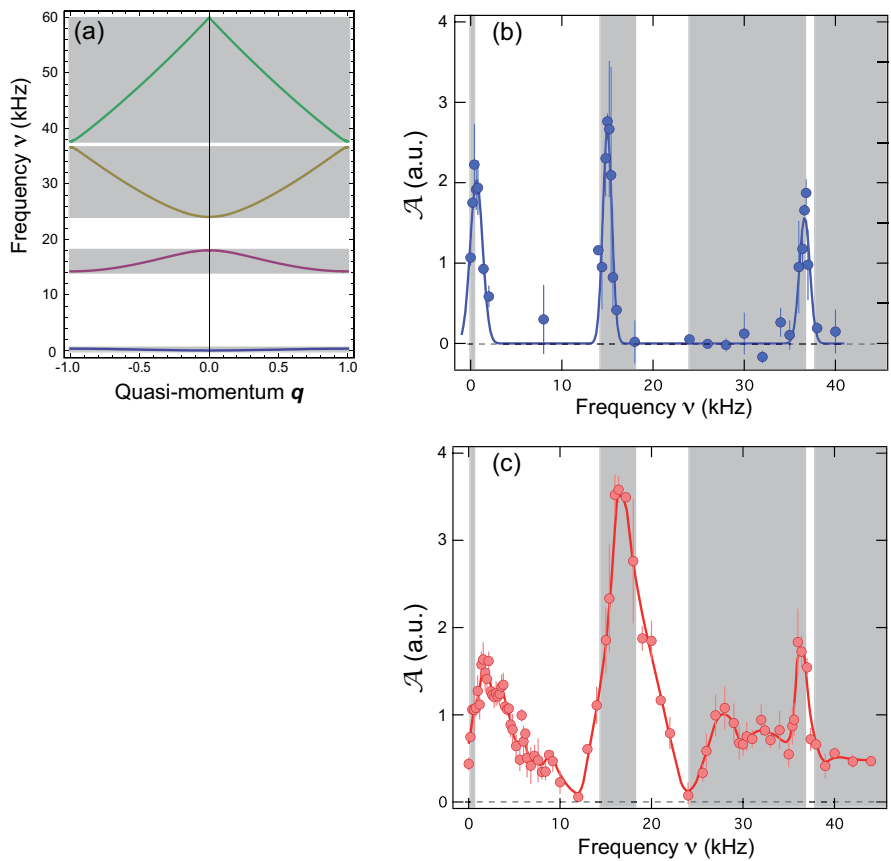}
\end{center}
\caption{(a) Energy bands of single particles in a periodic potential of amplitude $s=9$. The grey areas cover the entire energy distribution of each band and they are reported in Figs.~\ref{Fig5}(b)-(c). (b) Bragg spectrum of a 3D BEC loaded in a 1D optical lattice at the amplitude $s_{y}=9$ (blue dots) and for a momentum transfer $\hbar \mathrm{q}_{0,y}=0.96(3) \hbar k_{L}$. The solid blue line is a fit with three Gaussian functions. (c) Bragg spectra of an array of 1D BECs ($s_{\perp}=35$) loaded in a 1D optical lattice of amplitude $s_{y}=9$ (red dots), \emph{i.e.} the 1D gases being in the Mott-insulating state, and for a momentum transfer $\hbar \mathrm{q}_{0,y}=0.96(3) \hbar k_{L}$. The red solid line is a guide to the eye.}
\label{Fig5}
\end{figure}

As demonstrated for example in \cite{FabbriPRA2009}, the presence of a periodic potential implies the existence of different energy bands in the excitation spectrum, the so-called Bloch bands [see Fig.~\ref{Fig5}(a)]. The typical tunneling time of atoms in a lattice is $\sim \hbar/J$. When it is smaller than the time scale of the experiment, a 3D BEC loaded in the optical lattice has a long-range phase coherence. Its quasi-momentum distribution in the presence of the optical lattice has a small extension around the center of the first Brillouin zone $\mathrm{q}_{i,y}=0$. Therefore the resonance in the excitation spectrum is narrow since the resonant condition for a two-photon transition between momentum states $\hbar \mathrm{q}_{i,y}=0$ and $\hbar \mathrm{q}_{f,y}=\hbar \mathrm{q}_{0,y}$ is well defined in energy. Our experimental resolution is good enough to observe that this spread in energy is a small fraction of the energy bandwidth. As it is shown on Fig.~\ref{Fig5}(b), the experimental spectrum of this phase-coherent system ($s_{y}=9$, $s_{\perp}=0$) exhibits several well-defined resonances corresponding to excitations created in the different lattice bands \cite{FabbriPRA2009}.

The case of the array of 1D gases loaded in a 1D optical lattice is different, in particular when the amplitude of the longitudinal lattice is large enough for the 1D gases to be in the Mott-insulating regime. On the one hand, strong atom-atom correlations are present and are responsible for peculiar features of the Mott state such as the existence of an energy gap $\Delta_{\rm{ph}}$ in the excitation spectrum. On the other hand, the quasi-momentum distribution of a Mott insulator is spread over the first Brillouin zone ($-k_{L}<\mathrm{q}_{i,y}<k_{L}$) as a consequence of the spatial localization of atoms \cite{BlochRMP2008}. Deep enough in the Mott regime, this quasi-momentum distribution is expected to be close to homogeneous over the interval [$-k_{L};k_{L}$]. Therefore two-photon transitions towards the excited bands are possible from any initial quasi-momentum and Bragg spectroscopy can create excitations on a large energy interval, of the order of the energy bandwidth of single-particle. In this picture, fixing the frequency between the Bragg beams allows one to selectively excite (and therefore address) only a fraction of the atomic cloud through the resonance condition: for a given energy transfer lying in the energy interval of an excited band, one can always find a populated quasi-momentum state matching the resonance condition for being excited.

Fig.~\ref{Fig5}(c) depicts a spectrum of 1D gases in the Mott insulating state ($s_{\perp}=35,s_{y}=9$). The energy scale is identical to that of a 3D BEC loaded in a 1D optical lattice along the y-axis with the same amplitude $s_{y}=9$ [Fig.~\ref{Fig5}(b)]. On this energy scale, the spectra of 1D Mott insulating states exhibit several large resonances that have to be identified with transitions towards different energy bands of the optical lattice. The experimental measurements are reported with single-particle energy bands as grey areas. The lowest-energy band ($\nu < 10~$kHz) is much larger than the single-particle energy band due to the presence of atom-atom correlations. A detailed analysis of this effect has been the object of \cite{ClementPRL2009}. The energy transfer corresponding to transitions towards high energy bands is tens of kHz much larger than the Mott gap $\Delta_{\mathrm{ph}}\simeq h \times 2~$kHz \cite{ClementPRL2009}. Identifying the final momentum state of atoms giving contribution for the different parts of the excitation spectrum of Fig.~\ref{Fig5}c) should allow us to determine to which band the excitations belong to. To this purpose we have used a band mapping technic \cite{GreinerPRL2001} that we describe in the following paragraph.

\subsection{Measurement of the momentum of the excited atoms by the Bragg beams}

\subsubsection{Band population in a 1D optical lattice}

\begin{figure}[ht!]
\begin{center}
\includegraphics[width=0.7\columnwidth]{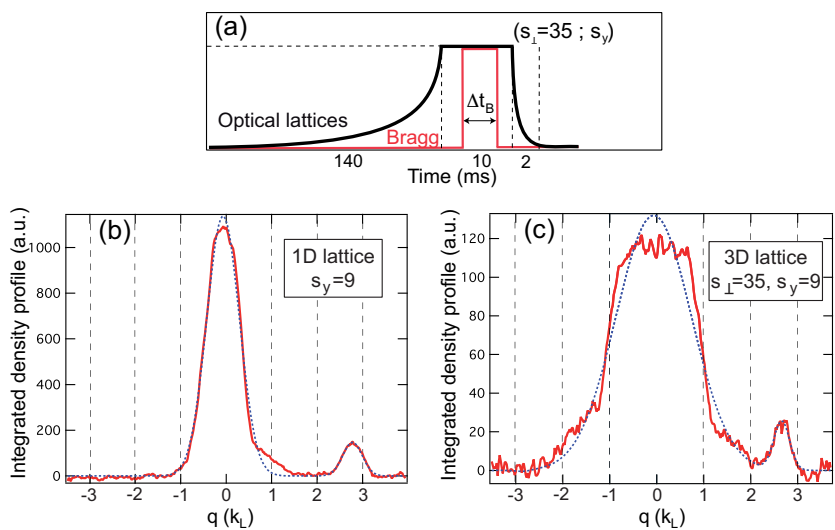}
\end{center}
\caption{(a) Experimental time sequence of the band-mapping technic. (b) Band population in the case of a 3D BEC in the presence of a 1D optical lattice of amplitude $s_{y}=9$ after a Bragg pulse with detuning $\nu=38~$kHz. (c) Band population in the case of an array of 1D BECs loaded in a 3D optical lattices of amplitudes $(s_{\perp}=35,s_{y}=9)$ after a Bragg pulse with detuning $\nu=34~$kHz. The horizontal scale is normalized to the momentum $\hbar k_{L}$ of the longitudinal optical lattice in both pictures.}
\label{Fig6}
\end{figure}

In order to identify the band towards which atoms are excited by the Bragg pulse we switch off abruptly the magnetic trap and we ramp down the lattice on a time scale ($\sim 2~$ms) too short for interaction-induced thermalization but adiabatic with respect to the atomic motion in a single lattice well \cite{GreinerPRL2001}. Doing this, one ensures to preserve the band population. After a time-of-flight, the atomic distribution reflects the band population in the momentum space. In practice, we ramp down the optical lattice with an exponential ramp of duration $2~$ms and time constant $0.5~$ms as sketched in Fig.~\ref{Fig6}(a). We first apply this technic to the 3D BEC loaded in a 1D optical lattice with amplitude $s_{y}=9$ corresponding to the spectrum of Fig.~\ref{Fig5}(b). The density profile along the y-axis after time-of-flight is shown on Fig.~\ref{Fig6}(b) when a resonant Bragg pulse creates excitations in the third energy band ($\nu =38~$kHz). The central peak around $\rm{q}=0$ corresponds to the non-excited cloud and it is the only feature left when the Bragg pulse is tuned out of resonance. On resonance a small lateral peak appears corresponding to excited atoms \cite{NoteExcitedAtoms}. The momentum of the excited atoms lies between $2 \hbar k_{L}$ and $3 \hbar k_{L}$, proving that these atoms are excited in the third band of the optical lattice. Fitting the position of the diffracted cloud with respect to the non-diffracted one with a Gaussian function we have measured momentum transfers of $1.1(1) \hbar k_{L}$ and $2.8(1) \hbar k_{L}$ for a Bragg pulse corresponding respectively to the transition towards the second and third energy band of the lattice. These results are in good agreement with the expected values $1.04(3) \hbar k_{L}$ and $2.96(3) \hbar k_{L}$ corresponding to a momentum transfer $\hbar \mathrm{q}_{0,y}=0.96(3) \hbar k_{L}$ given by the Bragg beams in our configuration \cite{NoteQuasiMom}.  

\subsubsection{Band population in the Mott insulating states}

We turn to the case of the 1D gases in the inhomogeneous MI state ($s_{\perp}=35,s_{y}=9$). The same experimental technic as described above is used to map the band population also in this case. The density profile obtained from the band mapping of the array of 1D gases is shown on Fig.~\ref{Fig6}(c). The detuning between the Bragg beams is $\nu=34~$kHz and corresponds to an excitation that lies in the wide and flat region of the spectrum. As in the case of Fig.~\ref{Fig6}(b), the diffracted atoms are clearly visible and lie in the third band of the optical lattice ($2 k_{L} < q < 3 k_{L}$). We note that the central (non-diffracted) peak is larger and flatter over the lowest lattice band with respect to the case of Fig.~\ref{Fig6}(b). This observation indicates that the quasi-momentum distribution of the inhomogeneous MI state covers almost uniformly the first band as expected for a system that is not phase-coherent. 

\begin{figure}[ht!]
\begin{center}
\includegraphics[width=.45\columnwidth]{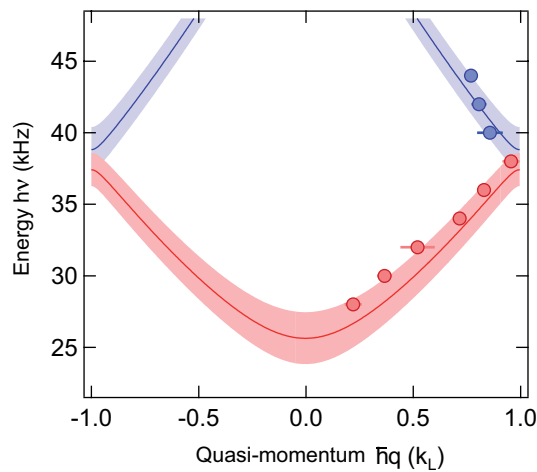}
\end{center}
\caption{Energy dispersion as a function of the quasi-momentum from the band-mapping technic used after a Bragg pulse. The dots correspond to the experimental measurements and the red and blue colors are associated respectively to excitations in the third and fourth band of the optical lattice. The solid lines are the band dispersion relation for single particles at $s_{y}=10$ and the shaded areas corresponds to the band dispersion relation taking into account the experimental uncertainty on the calibration of $s_{y}$ ($10 \%$).}
\label{Fig7}
\end{figure}

By applying the same fitting procedure as in the case of Fig.\ref{Fig6}(b), we extract the momentum $\hbar q$ to the diffracted atoms in the case of a 3D lattice with amplitudes ($s_{\perp}=35,s_{y}$). We repeat this procedure over the entire frequency range corresponding to the transition towards the third and fourth bands, \textit{i.e.} varying the relative detuning $\nu$ of the Bragg beams from $27$ to $45~$kHz, with the results plotted in Fig.~\ref{Fig7} for $s_{y}=10$. The experimental resolution $\simeq 400~$Hz of the Bragg spectroscopy allows us to do a precise mapping on this large energy scale. In Fig.~\ref{Fig7} we plot the energy transfer $h \nu$ given by the Bragg beams as a function of the momentum of the excited atoms measured using the band-mapping technic. These results are compared with the dispersion relation of single-particles in the presence of a periodic potential with an amplitude $s_{y}=10$ showing a good agreement \cite{EnergyOffset}. This demonstrates that: \emph{(i)} the excitations observed over a large energy scale between $27$ and $36~$kHz (\emph{resp.} between $37~$kHz and $45~$kHz) correspond to transition towards the third (\emph{resp.} fourth) energy band of the optical lattice ; \emph{(ii)}, the quasi-momentum distribution of the inhomogeneous MI state extends over the entire lowest lattice band since the entire energy band can be mapped using a Bragg excitation with a fixed momentum transfer $\hbar \rm{q}_{0,y}$. 

\subsection{High-energy bands, towards novel information about the Mott state}

Using a band-mapping technic we have demonstrated that the different parts of the spectrum measured in a Mott insulating state [see Fig.~\ref{Fig5}(c)] can be attributed to the different energy bands induced by the optical lattice. At high-energies, \emph{e.g.} for excitations into the third and the fourth bands, the frequency range of the Mott excitations corresponds to the bandwidth of the single-particle spectrum, a feature which is expected since in that situation the typical atom-atom correlation energy $\Delta_{\rm{ph}}$ is much smaller than that of the excited atoms. Yet, as we discuss it below,  the peculiar lineshape of the response within a given band has to be related to the properties of the many-body state which is probed. Therefore the identification of the energy scale of a given excited band opens the possibility to study the modulation of the response to the Bragg pulse \emph{within} a single band. For example, the response of the Mott insulating state to the Bragg spectroscopy shows a structure in amplitude within the third band (from $25$ to $36~$kHz) on Fig.~\ref{Fig5}(c). In particular, we observe an increased response at both edges of the band. A peak is mostly evident close to the higher-energy edge, the position of which corresponds to a Bragg transition from an initial quasi-momentum $q_{i}=0$. We now schematically discuss a way to interpret these features.

We expect the atoms excited by the Bragg beams in high-energy bands not to interact with the non-excited ones since their typical energy (corresponding to a frequency of several tens of kHz) is much larger than the particle-hole excitation energy $\Delta_{\mathrm{ph}}\simeq h \times 2~$kHz. In addition, as already mentioned, with our experimental parameters only a minor fraction of the atoms is excited. Therefore the transition induced by the Bragg beams in high-energy bands can be considered to happen between an initially strongly correlated state (Mott state) and single-particle states: the two-photon transition creates a hole in the Mott state and it populates highly energetic single-particle states. In other words, the response of the gas to the Bragg pulse gives access to the \emph{one-particle spectral function} of the Mott and it implies both the dispersion relation of the hole in the Mott phase and the density of states (DOS) to populate the excited band from the initial correlated phase \cite{DiscussAltman}. Information about the one-particle spectral function of solid-state systems has proved to be crucial to understand properties of correlated states such as high-Tc superconductors \cite{DamascelliRMP2003} and recent proposals have demonstrated the interest of measuring the one-particle spectral function for ultracold fermions \cite{DaoPRL2007}. We anticipate that the information about the one-particle spectral function obtained from the high-energy bands spectra might shed new light on the bosonic Mott state, in particular close to the SF-MI transition. To our knowledge, no theoretical predictions of such an experimental signal measured with ultracold bosons exist. 

The experimental situation is more complex than the case of a homogeneous Mott state. Due to the presence of the longitudinal trap, the 1D gases are in the inhomogeneous MI state formed of Mott regions with different fillings separated by superfluid areas \cite{BatrouniPRL2002}. This state can be viewed as constituted of domains with narrow quasi-momentum distributions (superfluid regions) and domains with almost homogeneous quasi-momentum distributions (MI regions), resulting in a non-trivial momentum distribution of the global state. As $U/J$ increases and the system is driven deeper in the Mott phase, the number of atoms which do not belong to a Mott state drops. For this reason we expect the contribution of the superfluid domains in the response to the Bragg excitation not to be significant. However, a complete picture taking into account contributions from the different Mott and superfluid regions has to be developed to interpret the details of the spectrum of Fig.~\ref{Fig5}c). We hope that the measurements presented in this paper could stimulate such theoretical work.
  
\section{Conclusion}

In conclusion, we have measured the response of bosonic Mott insulating states to excitations induced by inelastic scattering of light. As already demonstrated in \cite{ClementPRL2009}, this spectroscopic technic is a promising tool to characterize strongly interacting quantum phases. We have presented here in detail the experimental setup and the way the measurement has been carried out. In particular we have demonstrated that the quantity we measure is proportional to the increase of energy in the gaseous system and that the response to the light scattering is in a perturbative and linear regime. 

We have extended our previous work \cite{ClementPRL2009} to the study of excitations towards several energy bands in Mott insulating states. In contrast to the case of a 3D BEC loaded in a 1D optical lattice, the spectra of these Mott states exhibit broad resonances in energy. The use of a band-mapping technic after applying the Bragg pulse allows us to identify these resonances with transitions towards the different energy bands induced by the optical lattice. We also give direct experimental evidence that the momentum distribution of Mott states spreads over the entire first Brillouin zone. This property enables us to reconstruct the dispersion relation of the high energy bands using a Bragg excitation at a fixed momentum transfer. 

Finally the amplitude of the response of the inhomogeneous Mott insulating state in high energy bands exhibits peculiar structures within a single excited band. We expect that these could give important information on the experimental system and we think that this point should be the object of future investigation, both experimental and theoretical.

\section{Acknowledgments}

We acknowledge E. Altman, S. Huber and all the colleagues of the Quantum Degenerate Group at LENS for stimulating discussions. This work has been supported by UE contract No. RII3-CT-2003-506350, MIUR PRIN 2007, Ente Cassa di Risparmio di Firenze, DQS EuroQUAM Project, NAMEQUAM project and Integrated Project SCALA. This research was supported by a Marie Curie Intra European Fellowship within the 7th European Community Framework Programme (D.C.).


\section{References}

\begin{thebibliography}{99}

\bibitem{DagottoRMP1994} Dagotto~E 1994 {\it Rev. Mod. Phys.} {\bf 66} 763

\bibitem{NoziereBook} Nozieres~P and Pines~D 1994 {\it The Theory of Quantum Liquids} (Addison-Wesley, Reading)

\bibitem{GiamarchiBook} Giamarchi~T 2004 {\it Quantum Physics in One Dimension} (Oxford Science Publications, Oxford)

\bibitem{IshiiNature2003} Ishii~H \etal, 2003 {\it Nature (London)} {\bf 426} 540

\bibitem{AuslaenderScience2002}  Auslaender~O~M, Yacoby~A, de Picciotto~R, Baldwin~K~W,
Pfeiffer~L~N and West~K~W 2002 {\it Science} {\bf 295} 825

\bibitem{SchwartzPRB1998} Schwartz~A, Dressel~M, Gr\"uner~G, Vescoli~V, Degiorgi~L and Giamarchi~T 1998 {\it Phys. Rev. B} {\bf 58} 1261

\bibitem{StoferlePRL2004} St\"oferle~T, Moritz~H, Schori~C, K\"ohl~M and Esslinger~T  2004 {\it Phys. Rev. Lett.} {\bf92} 130403

\bibitem{SpielmanPRL2007} Spielman~I~B, Phillips~W~D and Porto~J~V 2007 {\it Phys. Rev. Lett.} {\bf 98} 080404

\bibitem{GreinerNature2002} Greiner~M, Mandel~O, Esslinger~T, H\"ansch~T~W and Bloch~I 2002 {\it Nature (London)} {\bf415} 39

\bibitem{FallaniPRL2007} Fallani~L, Lye~J~E, Guarrera~V, Fort~C and Inguscio~M 2007 {\it Phys. Rev. Lett.} {\bf 98} 130404

\bibitem{JordensNature2008} J\"ordens~R, Strohmaier~N, G\"unter~K, Moritz~H and Esslinger~T 2008 {\it Nature (London)} \textbf{455} 204

\bibitem{SchneiderScience2008} Schneider~U, Hackerm\"uller~L, Will~S, Best~T, Bloch~I, Costi~T~A,
Helmes~R~W, Rasch~W and Rosch~A 2008 {\it Science} \textbf{322} 1520

\bibitem{LignierPRA2009} Lignier~H, Zenesini~A, Ciampini~D, Morsch~O, Arimondo~E, Montangero~S, Pupillo~G and Fazio~R (2009) {\it Phys. Rev. A} \textbf{79} 041601(R)

\bibitem{FollingNature2005} F\"olling~S, Gerbier~F, Widera~A, Mandel~O,
Gericke~T and Bloch~I 2005 {\it Nature (London)} \textbf{434} 481

\bibitem{GuarreraPRL2008} Guarrera~V, Fabbri~N, Fallani~L, Fort~C, van der Stam~K~M~R and Inguscio~M 2008 {\it Phys. Rev. Lett.} {\bf 100} 250403

\bibitem{OvchinnikovPRL1999} Ovchinnikov~Yu~B, M\"uller~J~H, Doery~M~R, Vredenbregt~E~J~D, Helmerson~K, Rolston~S~L and Phillips~W~D 1999 {\it Phys. Rev. Lett} {\bf 83} 284

\bibitem{JakschPRL1998} Jaksch~D, Bruder~C, Cirac~J~I, Gardiner~C~W and Zoller~P 1998 {\it Phys. Rev. Lett.} \textbf{81} 3108

\bibitem{BatrouniPRL2002} Batrouni~G~G, Rousseau~V, Scalettar~R~T, Rigol~M, Muramatsu~A, Denteneer~P~J and Troyer~M 2002 {\it Phys. Rev. Lett.} \textbf{89} 117203

\bibitem{CampbellScience2006} Campbell~G~K, Mun~J, Boyd~M, Medley~P, Leanhardt~A~E, Marcassa~L~G, Pritchard~D~E, Ketterle~W 2006 {\it Science} \textbf{313} 649-652

\bibitem{FollingPRL2006} F\"olling~S, Widera~A, Mu\"ller~T, Gerbier~F and Bloch~I 2006 {\it Phys. Rev. Lett.} \textbf{97} 060403

\bibitem{ClementPRL2009} Cl\'ement~D, Fabbri~N, Fallani~L, Fort~C and Inguscio~M 2009 {\it Phys. Rev. Lett.} {\bf 102} 155301

\bibitem{OostenPRA2005} van Oosten~D, Dickerscheid~D~B~M, Farid~B, van der Straten~P and Stoof~H~T~C 2005 {\it Phys. Rev. A} \textbf{71} 021601(R)

\bibitem{ReyPRA2005} Rey~A~M, Blakie~P~B, Pupillo~G, Williams~C~J and Clark~C~W 2005 {\it Phys. Rev. A} \textbf{72} 023407

\bibitem{PupilloPRA2006}  Pupillo~G, Rey~A~M and Batrouni~G~G 2006 {\it Phys. Rev. A} \textbf{74} 013601

\bibitem{KollathPRL2006} Kollath~C, Iucci~A, Giamarchi~T, Hofstetter~W and Schollw\"ock~U 2006 {\it Phys. Rev. Lett.} \textbf{97} 050402

\bibitem{HuberPRB2007} Huber~S~D, Altman~E, B\"uchler~H~P and Blatter~G 2007 {\it Phys. Rev. B} \textbf{75} 085106

\bibitem{MenottiPRB2008} Menotti~C and Trivedi~T 2008 {\it Phys. Rev. B} \textbf{77} 235120

\bibitem{VanHovePR1954} Van Hove~L 1954 {\it Phys. Rev.} \textbf{95} 249 - 262 (1954)

\bibitem{AshcroftBook} Ashcroft~N~W and Mermin~N~D 1976 {\it Solid State Physics} (Saunders)

\bibitem{KozumaPRL1999} Kozuma~M, Deng~L, Hagley~E~W, Wen~J, Lutwak~R, Helmerson~K, Rolston~S~L and Phillips~W~D 1999 {\it Phys. Rev. Lett.} {\bf 82} 871

\bibitem{StengerPRL1999} Stenger~J, Inouye~S, Chikkatur~A~P, Stamper-Kurn~D~M, Pritchard~D~E and Ketterle~W 1999 {\it Phys. Rev. Lett.} \textbf{82} 4569

\bibitem{DavidsonRMP} Ozeri~R, Katz~N, Steinhauer~J and Davidson~N 2005 {\it Rev. Mod. Phys.} \textbf{77} 187

\bibitem{StamperKurnPRL1999} Stamper-Kurn~D~M, Chikkatur~A~P, G\"orlitz~A, Inouye~S, Gupta~S, Pritchard~D~E and Ketterle~W 1999 {\it Phys. Rev. Lett.} {\bf 83} 2876

\bibitem{Steinhauer2002} Steinhauer~J, Ozeri~R, Katz~N and Davidson~N 2002 {\it Phys. Rev. Lett.} \textbf{88} 120407

\bibitem{Steinhauer2003} Steinhauer~J, Katz~N, Ozeri~R, Davidson~N, Tozzo~C and Dalfovo~F 2003 {\it Phys. Rev. Lett.} \textbf{90} 060404

\bibitem{Richard2003} Richard~S, Gerbier~F, Thywissen~J~H, Hugbart~M, Bouyer~P and Aspect~A 2003 {\it Phys. Rev. Lett.} \textbf{91} 010405

\bibitem{Muniz2006} Muniz~S~R, Naik~D~S and Raman~C 2006 {\it Phys. Rev. A} \textbf{73} 041605

\bibitem{MullerPRL2008} See for example M\"uller~H, Chiow~S, Long~Q, Herrmann~S and Chu~S 2008 {\it Phys. Rev. Lett.} {\bf 100} 180405 and references therein

\bibitem{GerbierPRA2004} Gerbier~F, Thywissen~J~H, Richard~S, Hugbart~M, Bouyer~P and Aspect~A 2004 {\it Phys. Rev. A} {\bf 70} 013607

\bibitem{InadaPRL2008} Inada~Y, Horikoshi~M, Nakajima~S, Kuwata-Gonokami~M, Ueda~M and Mukaiyama~T 2008 {\it Phys. Rev. Lett.} {\bf 101} 180406

\bibitem{Papp2008} Papp~S~B, Pino~J~M, Wild~R~J, Ronen~S, Wieman~C~E, Jin~D~S and Cornell~E~A 2008 {\it Phys. Rev. Lett.} {\bf 101} 135301

\bibitem{Veeravalli2008} Veeravalli~G, Kuhnle~E, Dyke~P, Vale~C~J 2008 {\it Phys. Rev. Lett.} {\bf 101} 250403

\bibitem{GreinerPRL2001} Greiner~M, Bloch~I, Mandel~O, H\"ansch~T~W and Esslinger~T 2001 {\it Phys. Rev. Lett.} {\bf 87} 160405

\bibitem{DalfovoRMP1999} Dalfovo~F, Giorgini~S, Pitaevskii~L and Stringari~S 1999 {\it Rev. Mod. Phys.} {\bf 71} 463

\bibitem{CohenBook} Diu Laloe Cohen-Tannoudji 1977 Quantum Mechanics

\bibitem{MenottiPRA2002} Menotti~C and Stringari~S 2002 {\it Phys. Rev. A} {\bf 66} 043610

\bibitem{MoritzPRL2003} Moritz~H, St\"oferle~T, K\"ohl~M and Esslinger~T 2003 {\it Phys. Rev. Lett.} \textbf{91} 250402

\bibitem{MullerPRL2007} M\"uller~T, F\"olling~S, Widera~A and Bloch~I 2007 {\it Phys. Rev. Lett.} {\bf 99} 200405

\bibitem{NoteScaling} In \cite{ClementPRL2009}, we have used a slightly different scaling for the amplitude of the response with the parameters of the Bragg in order to normalize the different spectra, namely $(P_{B}^2 t_{B}/\delta^2)^{-1}$.

\bibitem{FabbriPRA2009} Fabbri~N, Cl\'ement~D, Fallani~L, Fort~C, Modugno~M, van der Stam~K~M~R and Inguscio~M 2009 {\it Phys. Rev. A} {\bf 79}  043623

\bibitem{ZambelliPRA2000} Zambelli~F, Pitaevskii~L, Stamper-Kurn~D~M and Stringari~S 2000 {\it Phys. Rev. A} {\bf 61} 063608

\bibitem{BlochRMP2008} Bloch~I, Dalibard~J and Zwerger~W 2008 {\it Rev. Mod. Phys.} \textbf{80} 885

\bibitem{NoteExcitedAtoms} To measure the spectra we excite fewer atoms than the ones which are seen after the band mapping technic. Indeed we have changed the parameters of the Bragg beams in the latter case in order to identify better the cloud of the excited atoms. 

\bibitem{NoteQuasiMom} Note that this technic is not well adapted to the measurement of the momentum transfer for the transition within the lowest energy band since the finite width of the non-diffracted atoms covers a large part of the first band.

\bibitem{SpielmanPRL2008} Spielman~I~B, Phillips~W~D and Porto~J~V 2008 {\it Phys. Rev. Lett.} \textbf{100} 120402

\bibitem{KinoshitaScience2004} Kinoshita~T, Wenger~T and Weiss~D 2004 {\it Science} \textbf{305} 1125

\bibitem{EnergyOffset} The dispersion relation in the lowest energy band can be considered as flat on the energy scale of Fig.~\ref{Fig7}. In Fig.~\ref{Fig7} we have substracted this constant energy offset to the bare calculation of the single-particle dispersion relation since it does not enter the two-photon transition process.

\bibitem{DiscussAltman} We acknowledge interesting discussions with E. Altman and S. Huber on this point.

\bibitem{DamascelliRMP2003} Damascelli~A, Hussain~Z and Shen~Z 2003 {\it Rev. Mod. Phys.} \textbf{75}, 473 

\bibitem{DaoPRL2007} Dao~T, Georges~A, Dalibard~J, Salomon~C and Carusotto~I 2007 {\it Phys. Rev. Lett.} \textbf{98} 240402

\end{thebibliography}
\end{document}